\documentclass{ws-ijmpa}

\begin{document}

\markboth{Yi-Fang Wang} {The Construction of the BESIII
Experiment}

%%%%%%%%%%%%%%%%%%%%% Publisher's Area please ignore %%%%%%%%%%%%%%%
%
\catchline{}{}{}{}{}
%
%%%%%%%%%%%%%%%%%%%%%%%%%%%%%%%%%%%%%%%%%%%%%%%%%%%%%%%%%%%%%%%%%%%%

\title{The Construction of the BESIII Experiment\footnote{ For the details
of the BESIII experiment and a full list of collaborators, please see http://bes.ihep.ac.cn.}
}

\author{Yi-Fang Wang}

\address{Institute of high energy physics, Beijing, 100049, P.R.
China\\
yfwang@mail.ihep.ac.cn}

\maketitle

\begin{history}
\received{Day Month Year}
\revised{Day Month Year}
\end{history}

\begin{abstract}
BESIII is a high precision, general purpose detector for the high
luminosity $e^+e^-$ collider, BEPCII, running at the tau-charm
energy region. Its design and current status of construction is
presented.

\end{abstract}

\ccode{PACS numbers: 11.25.Hf, 123.1K}

\section{Introduction}

The BESIII detector~\cite{besiii} is designed for the $e^{+}e^{-}$
collider running at the tau-charm energy region, called BEPCII,
which is currently under construction at IHEP, Beijing, P.R.
China. The accelerator has two storage rings with a circumstance
of 224~m, one for electron and one for positron, each with 93
bunches spaced by 8~ns~\cite{bepcii}.  The total current of the
beam is 0.93~amp, and the crossing angle of two beams is designed
to be 22~mrad.  The peak luminosity is expected to be $10^{33}
\mbox{cm}^{-2}\mbox{s}^{-1}$ at the beam energy of 1.89 GeV,  the
bunch length is estimated to be 1.5~cm and the energy spread will
be $5.16\times 10^{-4}$. At this moment, the LINAC has been
installed and successfully tested, all the specifications are satisfied.
The storage rings have installed, and will be commissioned for
synchrotron radiation run by the end of the year. The physics program
of the BESIII experiment includes light hadron spectroscopy, charmonium,
electroweak physics from charm mesons, QCD and hadron physics, tau
physics and search for new physics etc.  Due to its huge
luminosity and small energy spread, the expected event rate per
year is historical, as listed in table~\ref{tab:lum}.

\begin{figure}
\centerline{
\includegraphics*[width=80mm]{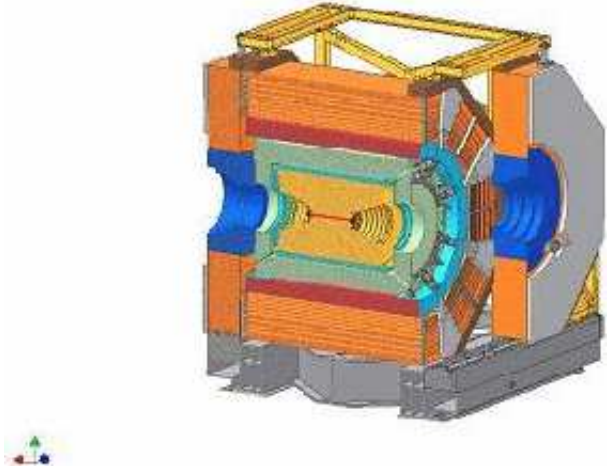}
}
 \caption{Schematics of the BESIII detector.}
 \label{fig:besiii}
\end{figure}

In order to achieve its physics goal and fully utilize the
potential of the accelerator, the BESIII detector~\cite{besiii},
as shown in Fig.~\ref{fig:besiii}, is designed to consist of a
drift chamber in a small cell structure filled with a helium-based
gas, an electromagnetic calorimeter made of CsI(Tl) crystals,
time-of flight counters for particle identification made of
plastic scintillators, a muon system  made of Resistive Plate
Chambers(RPC), and a super-conducting magnet providing a field of
1T. In the following, all the sub-detectors will be described
together with results of their performance tests.
\begin{table}[htbp]
\tbl{$\tau$-Charm productions at BEPC-II in one year's running($10^7s$).} {\begin{tabular}{@{}llll}
 \hline
               & Central-of-Mass energy & Luminosity     & \#Events  \\
Data Sample    & (MeV)& ($10^{33}$cm$^{-2}$s$^{-1}$)  & per year  \\
\hline
$J/\psi$ &  3097  & 0.6     & $10\times 10^9$\\
$\tau^+\tau^-$   & 3670 & 1.0  & $12\times 10^6$ \\
$\psi(2S)$ & 3686  & 1.0 & $3.0\times 10^9$ \\
$D^0\overline{D}^0$ & 3770 &1.0 & $18\times 10^6$ \\
$D^+D^-$ & 3770  &1.0 & $14\times 10^6$ \\
$D^+_S D^-_S$ & 4030  &0.6 & $1.0\times 10^6$ \\
$D^+_S D^-_S$ & 4170  &0.6 & $2.0\times 10^6$ \\
\hline
\end{tabular} \label{tab:lum}}
\end{table}

\section{Drift Chamber}

\begin{figure}
\centerline{
\includegraphics*[width=60mm]{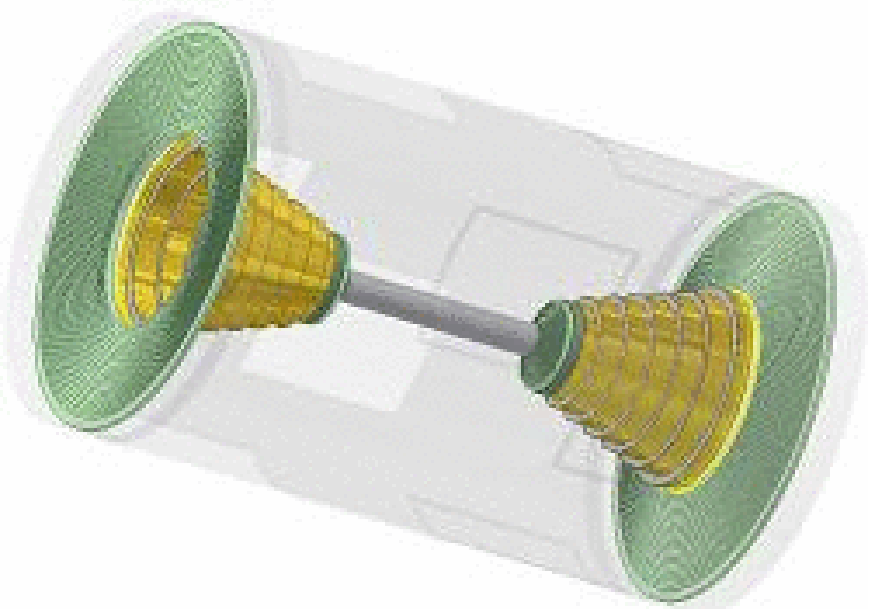}
\includegraphics*[width=60mm]{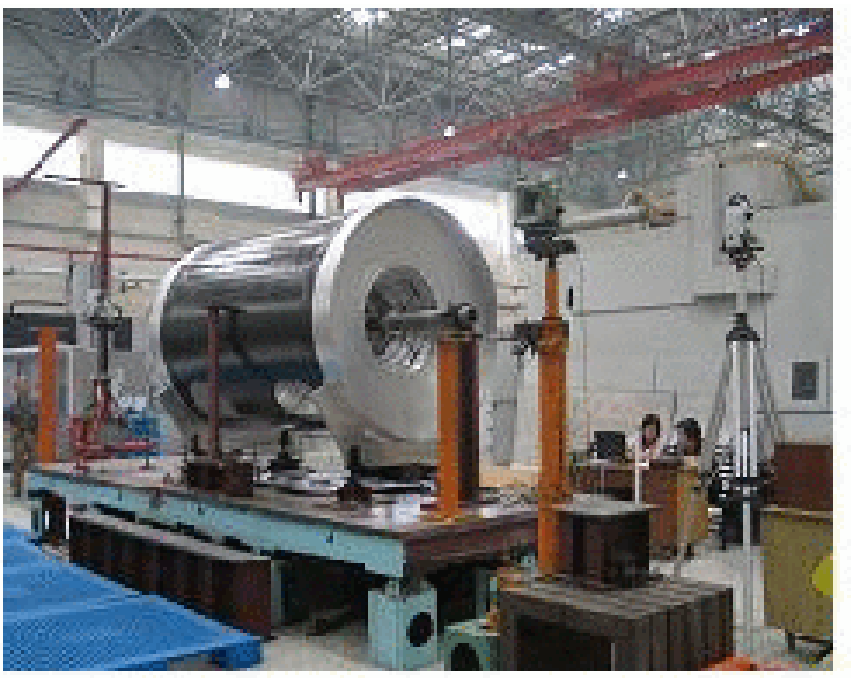}
}
 \caption{Left plot: the schematic view of the drift chamber. Right plot: the mechanical structure during the assembly.}
 \label{fig:mdc}
\end{figure}

The drift chamber has a cylindrical shape with two chambers
jointed at the end flange: an inner chamber without outer wall and
an outer chamber without inner wall. There are a total of 6
stepped end flanges made of 18 mm Al plates, as shown in left plot
of Fig.~\ref{fig:mdc}, in order to give space for the focusing
magnets. The inner radius of the chamber is 63 mm and the outer
radius is 810 mm, with a length of 2400 mm. Both the inner and
outer cylinder of the chamber are made of carbon fiber with a
thickness of 1 mm and 10 mm respectively. A total of 7000
gold-plated tungsten wires(3\% Rhenium) with a diameter of 25
$\mu$m are arranged in 43 layers, together with a total of 22000
gold-plated Al wires for field shaping. The small drift cell
structure of the inner chamber has a dimension of $6\times6
\mbox{mm}^2$ and the outer chamber of $8\times 8 \mbox{mm}^2$,
filled with a gas mixture of 60\% helium and 40\% propane.
 The designed single wire spatial resolution and dE/dX resolution are
130~$\mu$m and 6\%, respectively.

 The mechanical structure of the drift chamber, including the ultra-high precision(~20 $\mu$m)
 drilling of a total of ~30000 holes, and the  high
precision(~50 $\mu$m) assembly of 6 cylinders have been completed successfully.
 The right plot in Fig.~\ref{fig:assem} shows the mechanical structure during the assembly.
 The mass production of the feedthrough are completed and carefully tested one by one.
 A total of 30000 wiring are completed with a very high quality,
 the wire tension and the leakage current are well controlled.
 At this moment, the inner and outer chamber have been assembled together
 and the leak test for the helium gas is going on. The cosmic-ray test of the chamber in the laboratory will start soon.

\begin{figure}
\centerline{
\includegraphics*[width=60mm]{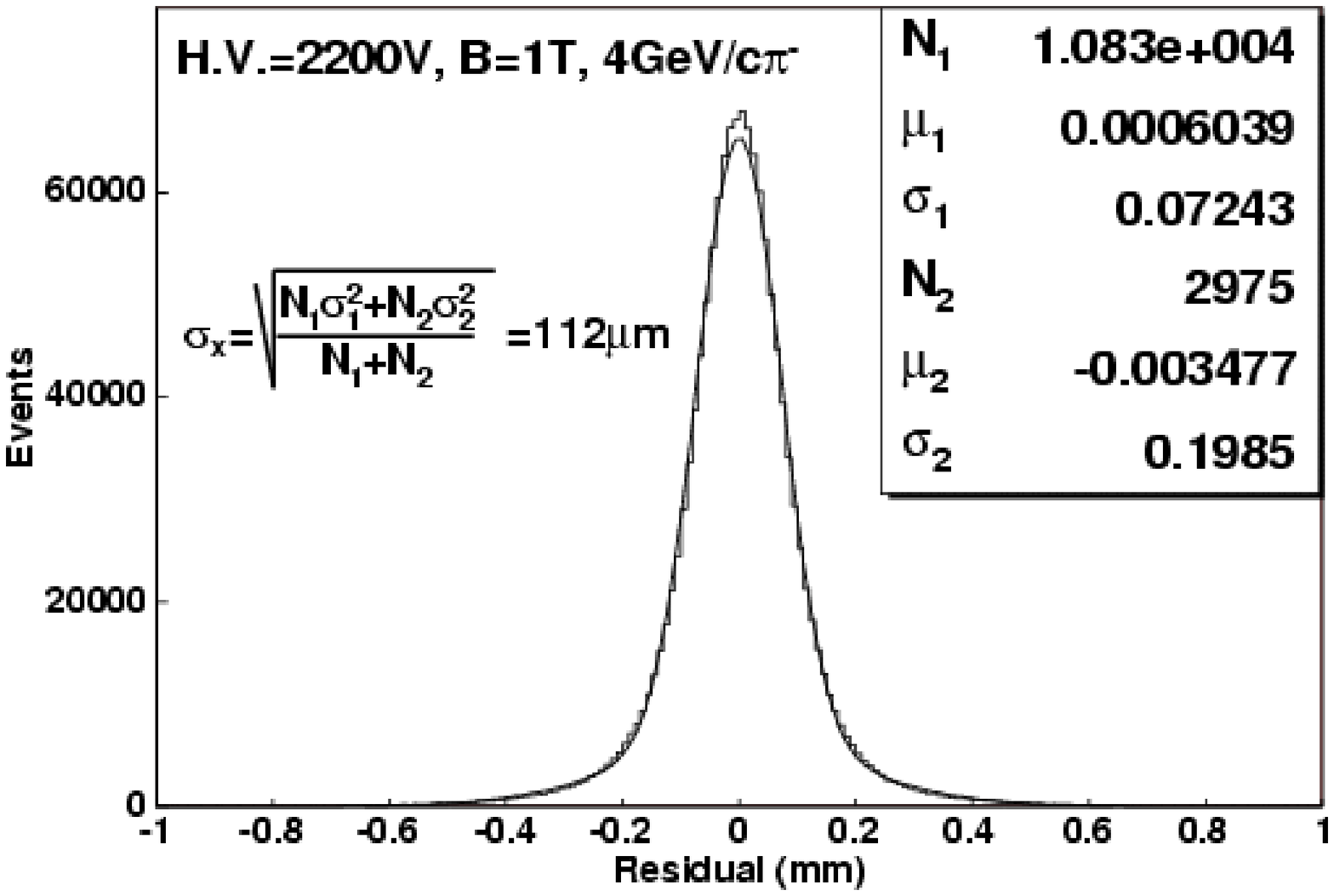}
\includegraphics*[width=60mm]{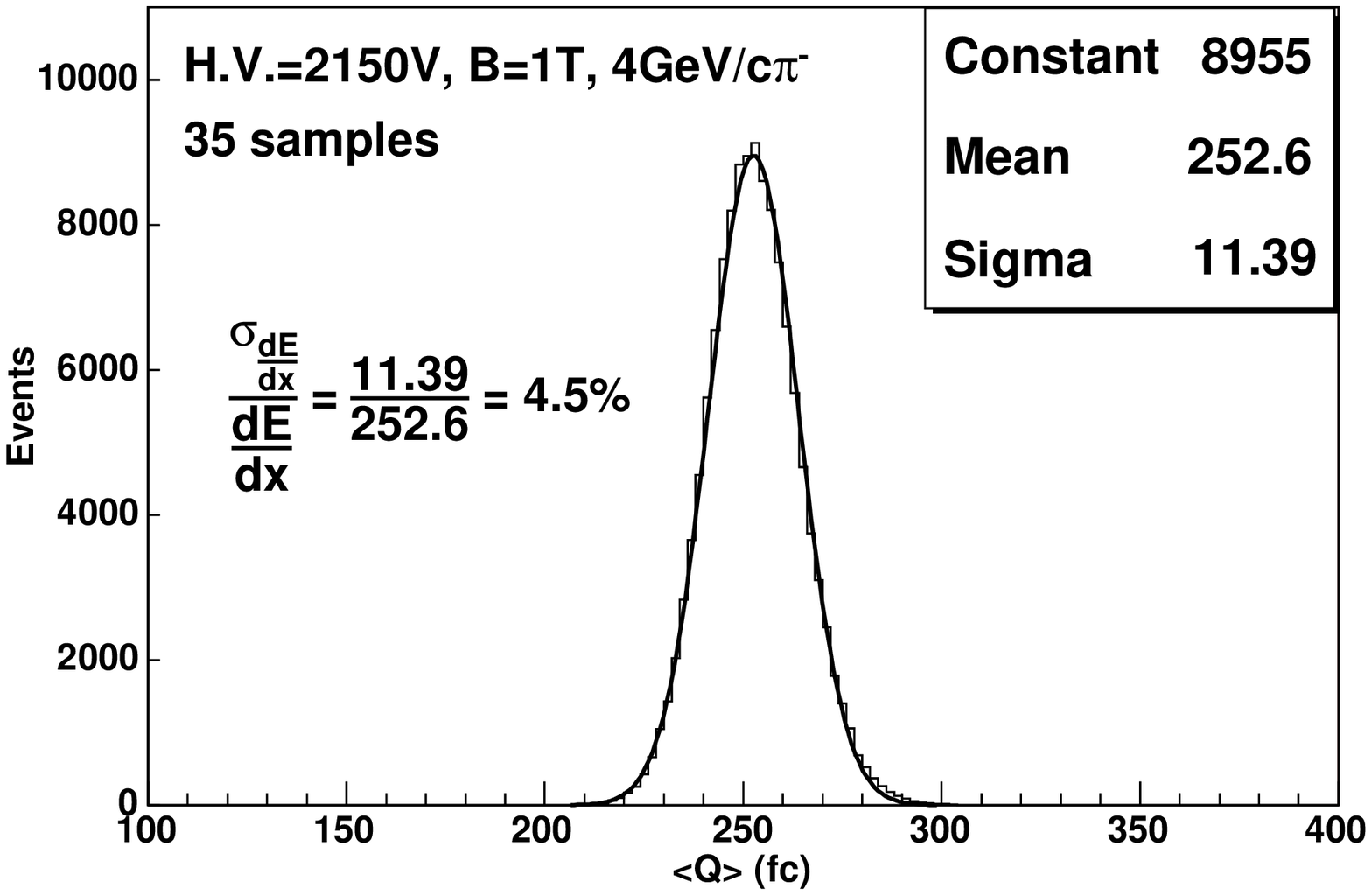}
}
 \caption{Left plot: the averaged single wire resolution.
 Right plot: the dE/dX resolution obtained from 80\% truncated mean.}
 \label{fig:reso}
\end{figure}

Several prototypes of the chamber have been tested at the beam in
KEK and IHEP\cite{jbliu,Qing}. Good results have been obtained in
all the cases. Fig.~\ref{fig:reso} shows a full length chamber
prototype tested in the IHEP E3  beam line using the actual setup of
160 channels of readout electronics, including amplifiers, readout
modules, cables, connectors, and the grounding setup. A prototype
of readout electronics with 512 channels including the data acquisition system
has been also tested in the laboratory for its long-term
stability. The mass production of all electronics boards is
almost finished, quality testing is underway.

\section{ CsI(Tl) Crystal Calorimeter}

\begin{figure}
\centerline{
\includegraphics*[width=80mm]{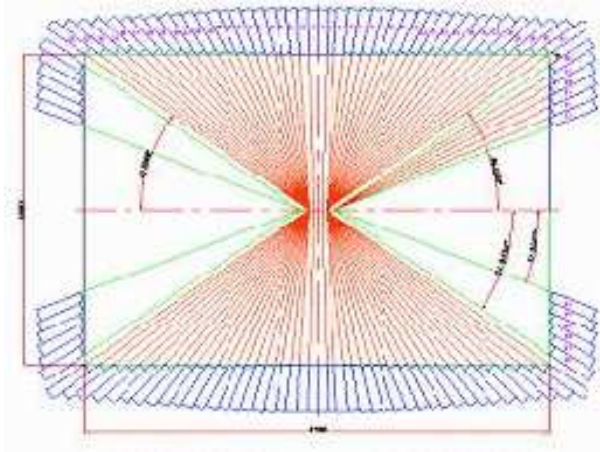}
}
 \caption{Schematic view of the CsI(Tl) crystal calorimeter.}
 \label{fig:csi}
\end{figure}

\begin{figure}
\centerline{
\includegraphics*[width=50mm]{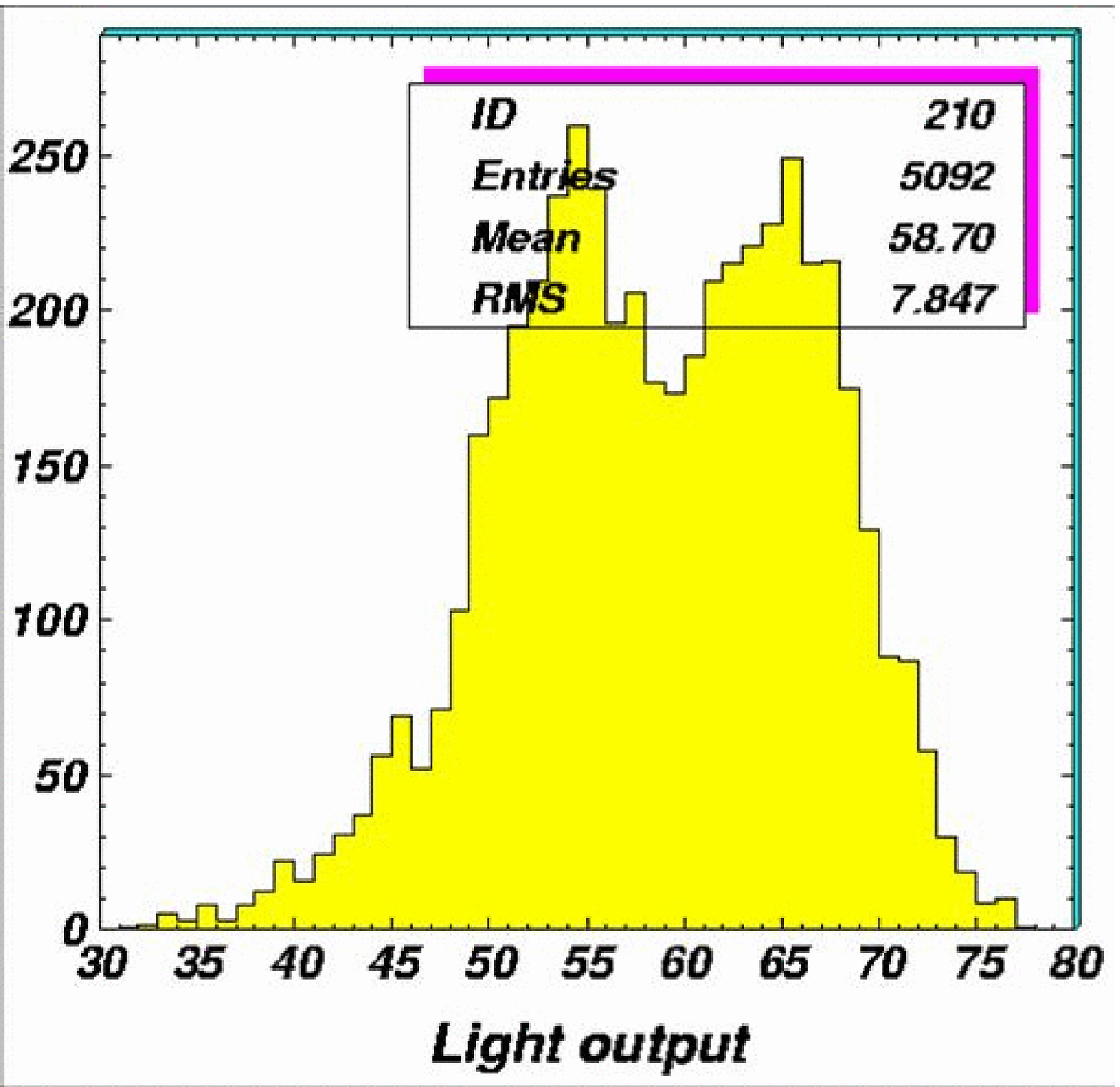}
\includegraphics*[width=50mm]{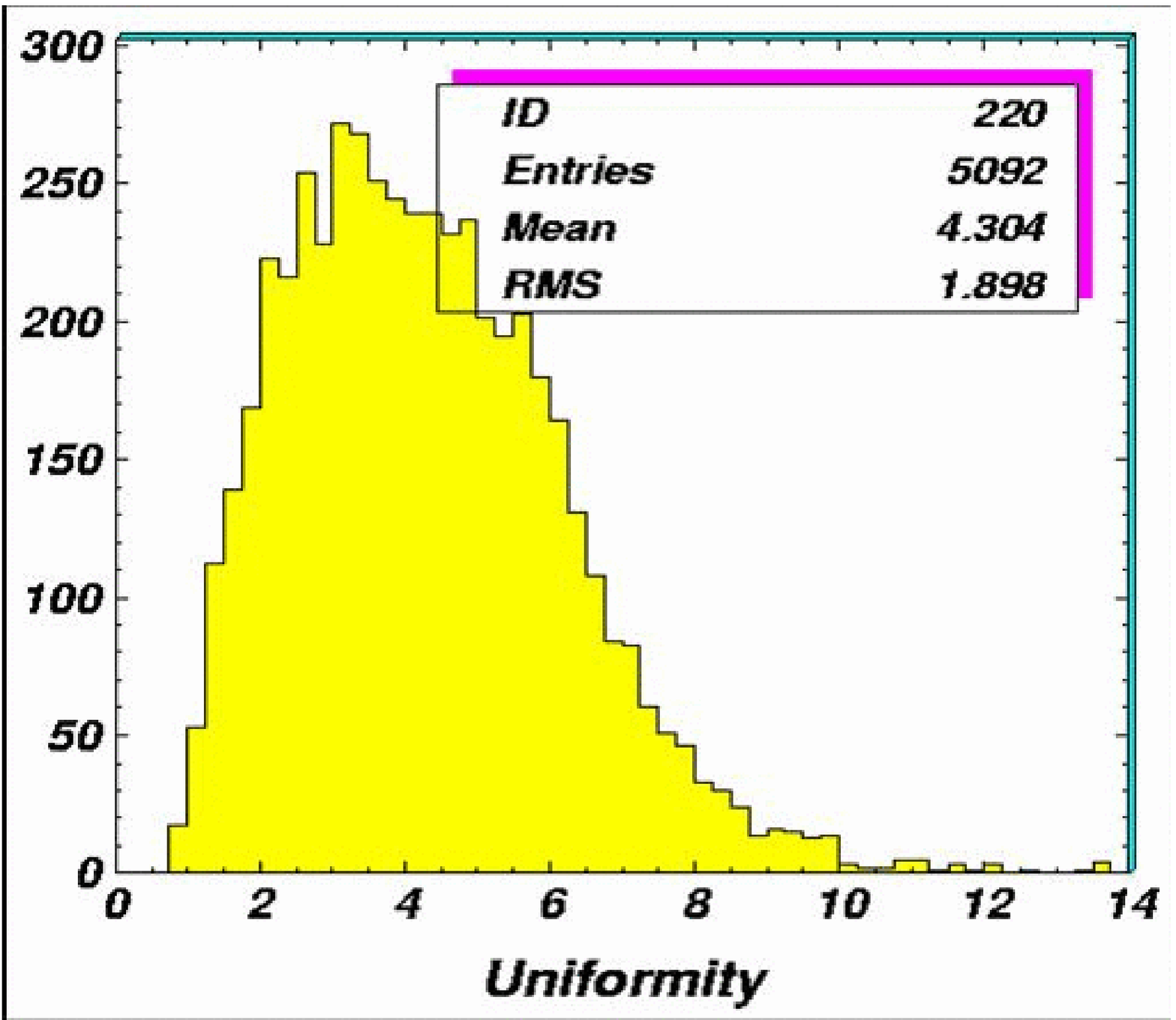}
}
 \caption{Left plot: The light yield of crystals; Right plot:
  The uniformity of crystals.}
 \label{fig:csi_performance}
\end{figure}

The CsI(Tl) crystal electromagnetic calorimeter consists of 6240
crystals, 5280 in the barrel, and 960 in two endcaps. Each crystal
is 28 cm long, with a front face of about $5.2\times 5.2
\mbox{cm}^2$, and a rear face of about $6.4\times 6.4
\mbox{cm}^2$. All crystals are tiled by 1.5~$^o$ in the azimuth
angle and 1-3~$^o$ in the polar angle, respectively, and point to
a position off from the interaction point by a few centimeters as
shown in Fig.~\ref{fig:csi}. They are hanged from the back by 4
screws without partition walls in order to reduce dead materials.
The designed energy and position resolution are 2.5\% and 6 mm at
1 GeV, respectively. At this moment all the barrel and half of the
endcap crystals have arrived, been tested, and assembled. The
light yield of arrived crystals is about 56\% with respect to the
reference crystal, as shown in left plot of
Fig.\ref{fig:csi_performance},
much more than the specification of more than 35%.
The average uniformity is better than 5\%, as shown in the right
plot of  Fig.~\ref{fig:csi_performance}, while the specification
is less than 7\%. All the photodiodes (PD) have been delivered, and their
performance before and after the accelerated aging, such as
dark current, noise, photon-electron
conversion efficiency and capacitance etc., have been tested.
All delivered crystals have been assembled and tested using cosmic rays.
The mechanical structure of the barrel is completed and the assembly of
crystals into the mechanical structure will start soon.


The readout electronics of crystals, including the preamplifier,
the main amplifier and charge measurement modules are tested at
the IHEP E3 beam line together with a crystal array and
photodiodes. Results from the beam test shows that the energy
resolution of the crystal array reached the design goal of 2.5\%
at 1 GeV and the equivalent noise achieved the level of less than
1000 electrons, corresponding to an energy of 220 keV. A prototype with
384 channels has been tested for long term stability. The mass
production of preamplifiers is finished, and other modules are
almost finished.

\section{Time-Of-Flight system}

The particle identification at BESIII is based on the momentum and
dE/dx measurements by the drift chamber, and the
Time-of-Flight(TOF) measurement by plastic scintillators. The
barrel scintillator bar is 2.4 m long, 5 cm thick and 6 cm wide. A
total of 176 such scintillator bars constitute two cylinders, to
have a good efficiency and time resolution. For the endcap, a
total of 48 fan-shaped scintillators form a single layer. A 2 inch
fine mesh phototube is directly attached to each scintillator to
collect the light. The intrinsic time resolution is designed to be
90 ps including contributions from electronics and the common time
corresponding to the beam crossing. Such a time resolution,
together with contributions from the beam size, momentum
uncertainty, etc. can distinguish charged $\pi$ from K mesons for
a momentum up to 0.9 GeV at the 2$\sigma$ level.
\begin{figure}
\centerline{
\includegraphics*[width=60mm]{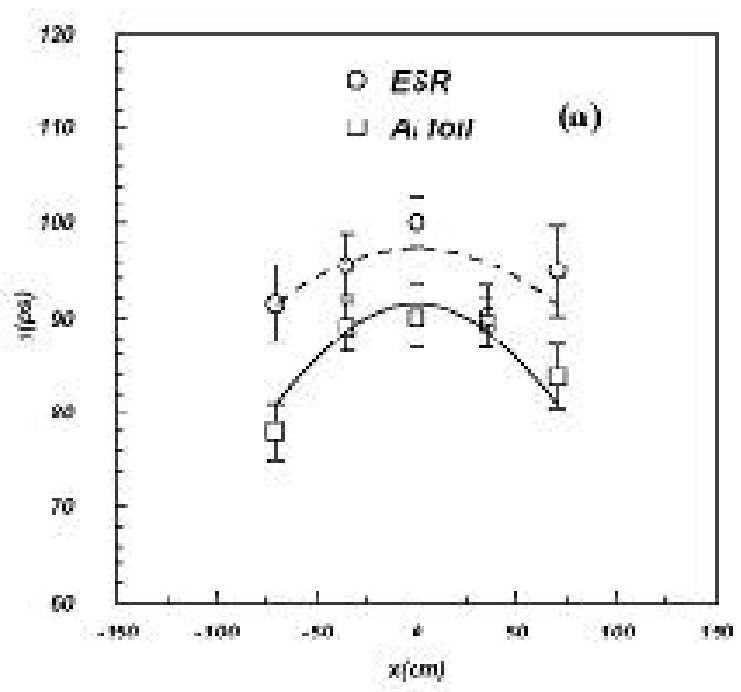}
\includegraphics*[width=60mm]{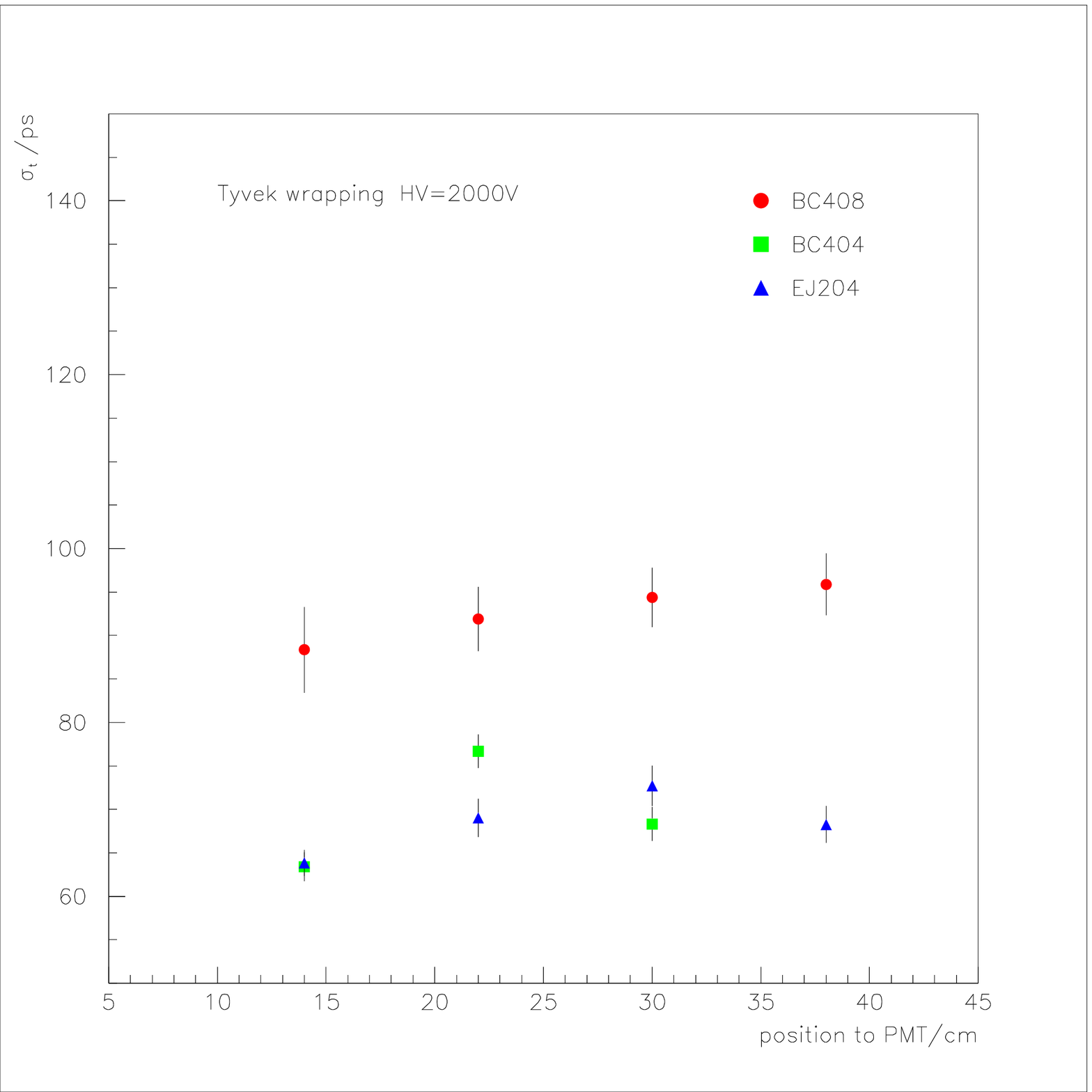}
}
 \caption{Left plot:Time resolution of the barrel
TOF module from a beam test. Right plot: Time resolution of the
endcap TOF module from a beam test.}
 \label{fig:tof_per}
\end{figure}

Beam tests of TOF prototypes have been performed at IHEP E3 beam
line using pions, electrons and protons~\cite{wu,an}. Different
scintillator types such as BC404, BC408 and EJ200, with different
thickness are tested, together with different wrapping materials.
The results, as shown in Fig.~\ref{fig:tof_per}, show that the
time resolution using a prototype of readout electronics including actual cables are better than 90 ps and 75 ps for the barrel  and the endcap, respectively.
Currently, most of the PMTs have been tested, and some of the
scintillators have been delivered. The mass production of
preamplifiers has been almost completed and other readout modules are about to start.

\section{Muon counter}

\begin{figure}
\centerline{
\includegraphics*[width=60mm]{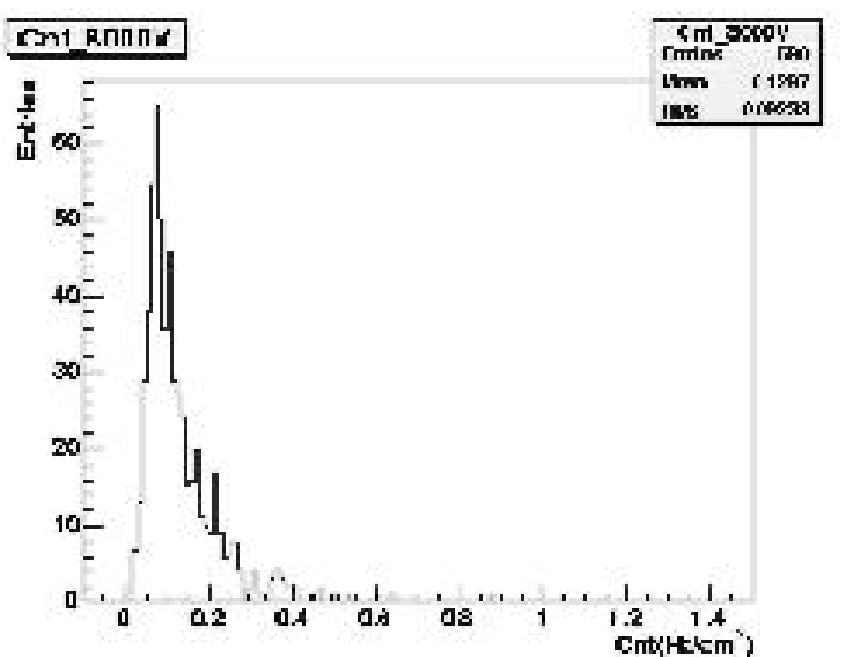}
\includegraphics*[width=60mm]{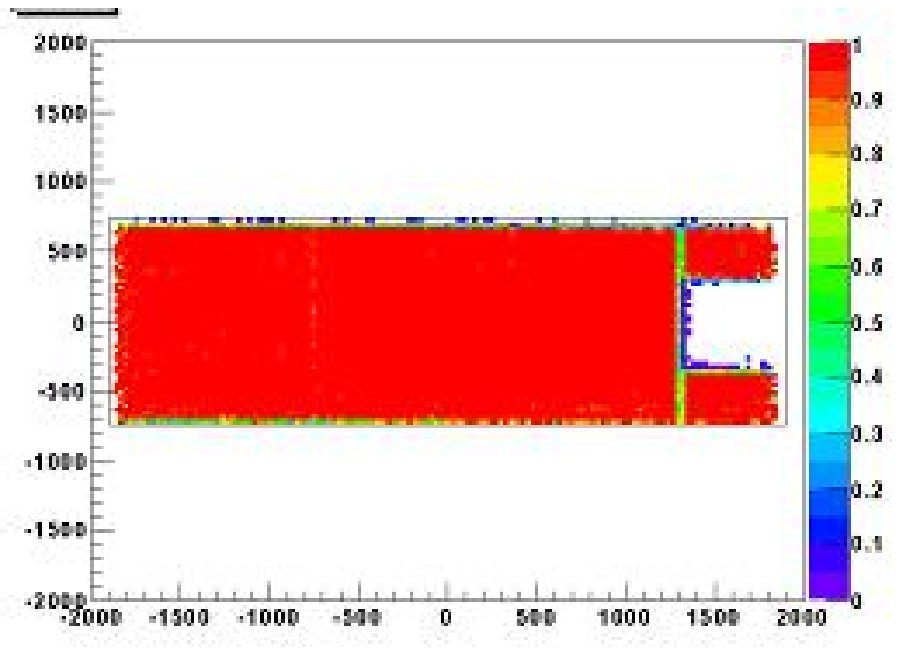}
}
 \caption{Left plot: Counting rate of RPC. Right plot:  Measured efficiency of a RPC after installation using cosmic rays.}
 \label{fig:rpc_count}
\end{figure}

\begin{figure}
\centerline{
\includegraphics*[width=80mm]{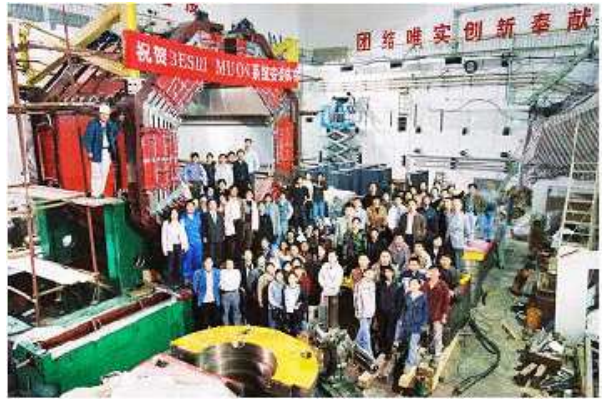}
}
 \caption{ A celebration for the successful installation of the magnet yoke interleaved with  Muon chambers made of RPC.}
 \label{fig:rpc_con}
\end{figure}

The BESIII muon chamber is made of Resistive Plate Chambers(RPC)
interleaved in the magnet yoke. There are a total of 9 layers in
the barrel and 8 layers in the endcap, with a total area of about
$2000$ m$^2$. The readout strip is 4 cm wide, alternated between
layers in x and y directions. The RPC is made of bakelite with a
special surface treatment without linseed oil~\cite{seri}.
Such a simple technique for the RPC production shows a good
quality and stability at a low cost.
 All RPCs have been manufactured, tested, assembled and installed with satisfaction.
 The left plot in Fig.~\ref{fig:rpc_count} shows the counting rate of all RPCs from the mass production after one week training.
Most of them has a noise rate of about 0.1 Hz/cm$^2$, which will be reduced to typically 0.04Hz/cm$^2$ after one month training.
 The right plot in Fig.~\ref{fig:rpc_count} shows the measured efficiency of installed RPCs, which is more than 95\% in all the region, using cosmic-rays.
The magnet yoke together with all RPC has been installed as shown
in Fig.~\ref{fig:rpc_con}.

\section{Superconducting magnet}
\begin{figure}
\centerline{
\includegraphics*[width=60mm]{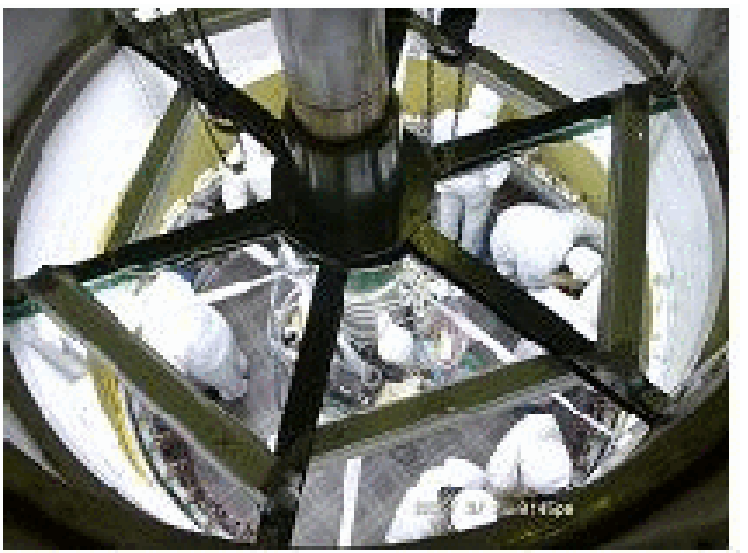}
\includegraphics*[width=60mm]{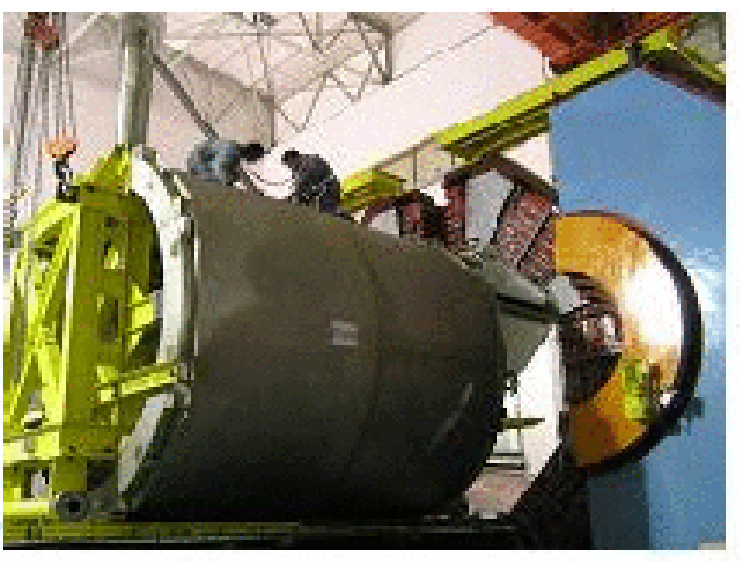}
}
 \caption{Left plot: Coil winding at IHEP. Right plot: the magnet during the installation.}
 \label{fig:sc_coil}
\end{figure}

The BESIII super-conducting magnet has a radius of 1.48 m and a
length of 3.52 m. It use the Al stabilized NbTi/Cu conductor with
a total of 920 turns, making a 1.0T magnetic field at a current of
3400 amp. The total cold mass is 3.6t with a material thickness of about
1.92 X$_0$. In collaboration with WANG NMR of California, the
magnet is designed and manufactured at IHEP.  The left plot in
Fig.~\ref{fig:sc_coil} shows the coil winding at IHEP by
technicians. The magnet was successfully installed into the iron
yoke of the BESIII, as shown in the right plot of
Fig.~\ref{fig:sc_coil}, together with the valve box. The magnet
has been successfully cool down to the super-conducting
temperature with a heat load within the specification.  A stable
magnetic field of 1.0T at a current of 3368 amp was achieved. The
dump resistor and dump diode, switches of the quench protection
devices are installed and tested successfully. The field mapping together with super-conducting quadrapole magnets for final focusing of the beam will start soon.

\section{Trigger and DAQ System}

\begin{figure}
\centerline{
\includegraphics*[width=90mm]{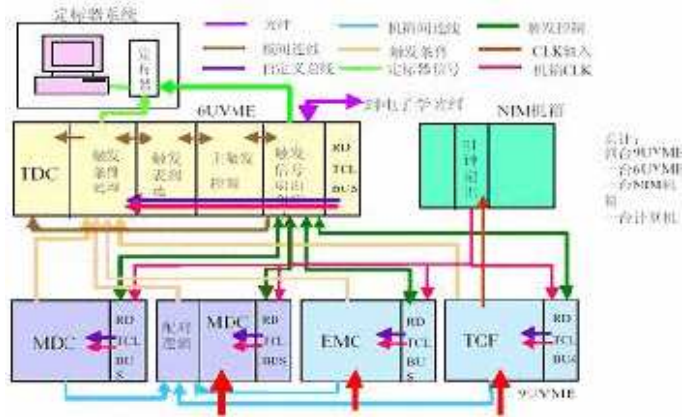}
}
 \caption{Trigger systems and their inter-connection.}
 \label{fig:trigger}
\end{figure}

The BESIII trigger rate is estimated to be about 4000 Hz and the
trigger system is designed largely based on the latest technology
such as fiber optics, pipelines and FPGA chips.
Fig.~\ref{fig:trigger} shows the schematics of the trigger systems
and their inter-connections. Information from sub-detector
electronics is feed into sub-detector trigger system via fiber
optical cables in order to avoid grounding loops. The VME based
main trigger and all the sub-trigger boards communicate with each
other via copper cables. All trigger logic stored in FPGA chips
are programmable and can be downloaded via VME bus. The trigger
latency is designed to be 6.4 $\mu$s and the pipeline technique is
used for all the readout electronics. The radiation hardness of
fiber cables and their connectors are tested at BEPC beam test
facility. Some of the sub-trigger systems share the same hardware
design of the board using different firmware in order to reduce
number of board types and save the cost. Latest large FPGA chips
with RocketIO technology are adopted in such a board design. All
the modules have been designed, prototyped, tested and some have
completed the mass production.

\begin{figure}
\centerline{
\includegraphics*[width=90mm]{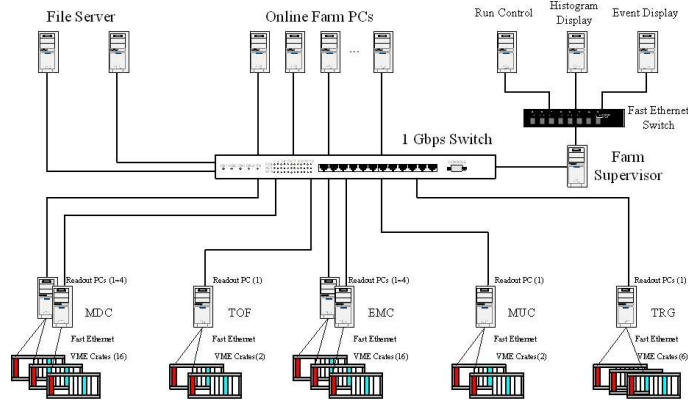}
}
 \caption{ The structure of the BESIII data acquisition system.}
 \label{fig:daq}
\end{figure}

The total data volume at BESIII is about 50 Mbytes/s for a trigger
rate of about 4000 Hz. The DAQ system shall read out the event
fragments from the front-end electronics distributed over more than
40 VME crates, and build them into a complete event to be
transmitted for recording on the persistent media. A simplified
structure of the BESIII DAQ system is shown in Fig.\ref{fig:daq}.
The DAQ software, based on the ATLAS TDAQ, includes database
configuration, data readout, event building and filtering,
run control, monitoring, status reporting and data
storage, etc. Every component has been tested
successfully at an average event rate of 8000 Hz and 4500 Hz with
an event size of 12KB and 25KB, respectively. The software has
been used for cosmic-rays and beam test for a Drift Chamber
prototype and an EMC crystal array. Different working modes such as
normal data taking, baseline, calibration, debugging of the readout
electronics and waveform sampling have been tested. As a
distributed system, the entire DAQ system must keep synchronized,
so a state machine is implemented in the PowerPC readout subsystem
to keep the absolute synchronization with the DAQ software, which
guarantees the coherency of the whole system.

\section{Offline computing and software}
\begin{figure}
\centerline{
\includegraphics*[width=60mm]{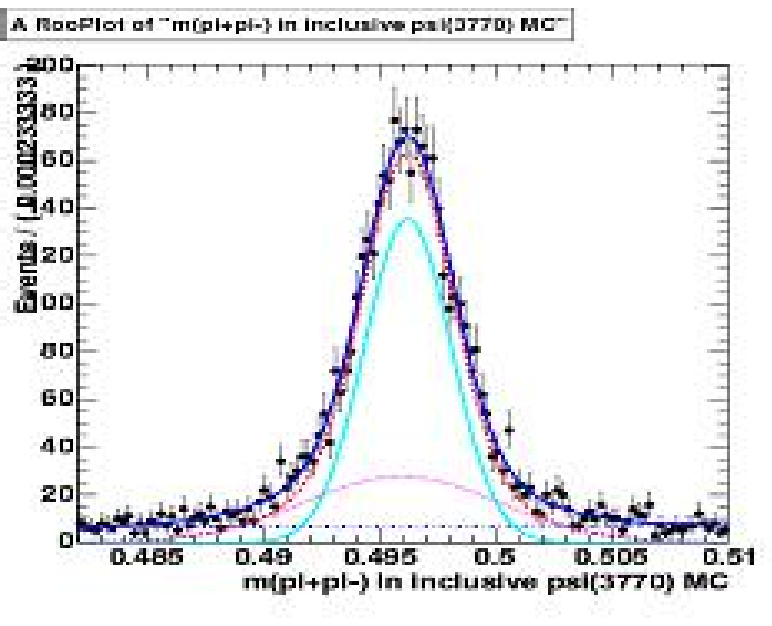}
\includegraphics*[width=60mm]{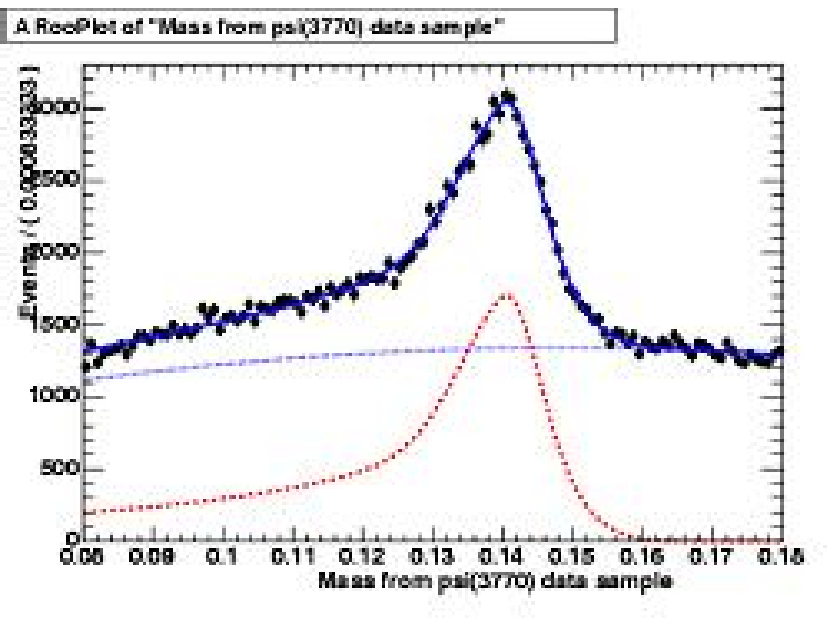}
}
 \caption{Left plot: the invariant mass of $\pi^+\pi^+$, the resolution of $K_S$ mass is about 3.0 MeV.
 Right plot: The reconstructed $\pi^0$ mass from $\gamma \gamma$ decay mode with resolution about 7.0 MeV.}
 \label{fig:offline}
\end{figure}

The BESIII offline computing system is designed to have a PC farm
of about 2000 nodes for both data and Monte Carlo production, as well as
data analysis. A computing center at IHEP and several local
centers at collaborating universities are anticipated. A 1/10
system will be built at IHEP by the end of the year and the full
system will be built next year.

The offline software consists of a
framework based on Gaudi, a Monte Carlo simulation based on
GEANT4, an event reconstruction package, a calibration and a database
package using MySQL. Currently all codes are working as a complete
system, and tests against cosmic-ray and beam test data are
underway. Analysis tools such as particle identification,
secondary vertex finding, kinematic fitting, event generator and
partial wave analysis are partially completed, although continuous
progress are expected. Figure~\ref{fig:offline} shows a
reconstructed $K_S$ and $\pi^0$ invariant mass with a resolution
as expected.

\section{Summary}

    The BEPCII and BESIII construction went on smoothly.
Currently all the R\&D programs have been completed successfully,
mass production of detector components is underway, and some
already assembled and installed. The detector installation is expected to
be completed by next year and the physics data taking will
start at the end of 2007.

\section*{Acknowledgments}

The author thanks all the BEPCII staff for their excellent work
towards the construction of the accelerator, and all the BESIII
staff for completing the detector R\&D and construction on time
and in good quality.

%\begin{thebibliography}{000} %for 3 digits
%\begin{thebibliography}{00}  %for 2 digits

\end{document}